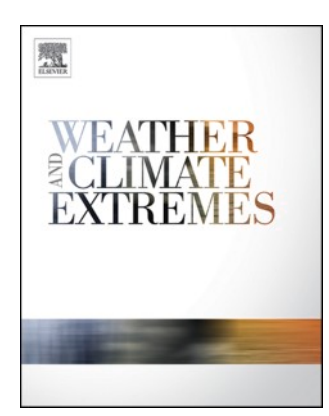

# Did recent sea surface temperature warming reinforce the extreme East Asian summer monsoon precipitation in 2020?☆

Taeho Mun [a], Haerin Park [a], Dong-Hyun Cha [a,*], Chang-Keun Song [a], Seung-Ki Min [b,c], Seok-Woo Son [d]

[a] Department of Civil, Urban, Earth and Environmental Engineering, Ulsan National Institute of Science and Technology, Ulsan, South Korea
[b] Division of Environmental Science and Engineering, Pohang University of Science and Technology, Pohang, South Korea
[c] Institute for Convergence Research and Education in Advanced Technology, Yonsei University, Seoul, South Korea
[d] School of Earth and Environmental Sciences, Seoul National University, Seoul, South Korea

A B S T R A C T

We analyzed the possible effects of recent sea surface temperature (SST) warming on the extraordinary East Asian summer monsoon (EASM) precipitation in 2020 summer. The dynamic and thermodynamic impacts of SST are examined by conducting regional climate model experiments with observed SST and cold SST where the 22-year SST trend is removed. In the presence of warm SST, precipitation increases in low latitudes but decreases in the EASM region. This dipolar precipitation change pattern opposes the precipitation anomalies in 2020 summer, indicating that the extraordinary 2020 EASM precipitation is not likely driven by recent SST warming. The warm SST suppresses the western North Pacific subtropical high expansion and weakens the southwesterly from the South China Sea toward the EASM region. In terms of large-scale atmospheric circulations, SST-induced wind changes strengthen the local Walker circulation in the South China Sea and the Philippines and the local Hadley circulation across the EASM region. These support the reduced EASM rainfall in the control experiment compared to the cold SST experiment and imply that the precipitation reduction by dynamical effects could exceed the precipitation increase by thermodynamic effects in the EASM region under warm SST.

## 1. Introduction

The Asian summer monsoon (ASM) is a vast and compound system that includes various temporal and spatial variabilities, and the East Asian summer monsoon (EASM) is an essential part of this system (Cha et al., 2011; Lau et al., 2000; Wang et al., 2021). In East Asia, at least 30–50% of the annual accumulated precipitation falls in summer, causing enormous social and economic damage owing to the variability of the EASM (Cha et al., 2008; Chen and Chang 1980; He and Zhou 2020; Tao and Chen 1987; Zhang et al., 2021). During the summer of 2020, the most overwhelming amount of precipitation was recorded in East Asia, including China, Korea, and Japan, from the 1980s until today (Fig. 1a, b, and e) (Takaya et al., 2020; Zhang et al., 2021). The Changma, the frontal rain band during the EASM season in Korea, lasted 54 days in South Korea, the longest since 1973 and the record-breaking longest rainy season in the country (KMA 2021). It caused flooding and landslides, resulting in 8.23 million victims including displaced people and casualties in China (Qian et al., 2022). The reason for the record-breaking rainfall was that the western North Pacific subtropical high (WNSPH) extended exceedingly to the northwest, creating an environment that could sustain fronts across China, Korea, and Japan and allow tropical cyclones to pass through the EASM region (Fig. 1c and d) (Park et al., 2021).

During the boreal summer, the expansion of the WNPSH supplies warm air and abundant moisture from the tropics to East Asia, forming a quasi-stationary front as it collides with cold and dry air to the north (Liu et al., 2020; Takaya et al., 2020). It is challenging to predict the variability of the EASM or to point out a specific phenomenon as the cause of the EASM because the factors of various temporal and spatial scales that affect the variability of the EASM interact with each other (Cha et al., 2011). Typically, the El Niño-Southern Oscillation (ENSO) has a significant influence on the interannual variability of the EASM, and the

☆ To be submitted to Weather and Climate Extremes.

* Corresponding author. Department of Civil, Urban, Earth and Environmental Engineering, Ulsan National Institute of Science and Technology, UNIST-gil 50, Ulsan, 44919, South Korea.
*E-mail address:* dhcha@unist.ac.kr (D.-H. Cha).

preceding El Niño results in anomalous expansion of the WNPSH, which induces a sufficient supply of moisture in East Asia (Xie et al., 2009; Zhang et al., 2021). The Indian summer monsoon and Western North Pacific Summer Monsoon (WNPSM) are also related to the intensity of the EASM precipitation because they affect moisture transport (Wang and LinHo 2002; Zhang et al., 1999). The summer North Atlantic Oscillation, the Madden Julian Oscillation (MJO), polar ice, Eurasian snow cover, land-surface processes, disturbances in the mid-latitudes, and thermodynamic and dynamic effects of the Tibetan Plateau may also affect the EASM variability (Cha et al., 2011; Huang et al., 2003; Liu et al., 2020).

There were weak previous El Niño events in the winter of 2020 (Cai et al., 2022). Pan et al. (2021) reported that fast decaying central Pacific El Niño and developing La Niña contributed to the enhanced WNPSH. However, the potential drivers of extreme EASM in 2020 remain to be investigated. Previous studies have suggested that sea surface temperature (SST) warming in the tropical North Indian Ocean, the Indo-western Pacific Ocean capacitor mode (IPOC), and long-lasting MJO induced the expansion of the WNPSH, leading to extreme precipitation over East Asia in the summer of 2020 (Guo et al., 2021; Park et al., 2021; Wang et al., 2021; Zhang et al., 2021; Zhou et al., 2021). The Arctic sea-ice loss and tropical North Atlantic warming were also suggested as possible factors of anomalous circulation and precipitation in 2020 EASM (Chen et al., 2022; Kim et al., 2022). Takaya et al. (2020) reported that the Rossby wave generated in the tropical region due to the strong positive Indian Ocean Dipole in 2019 reached the east coast of Africa in 2019/2020 winter and then propagated to the east as a Kelvin wave, which was similar to the mechanism occurring after the strong El Niño preceding it in East Asia.

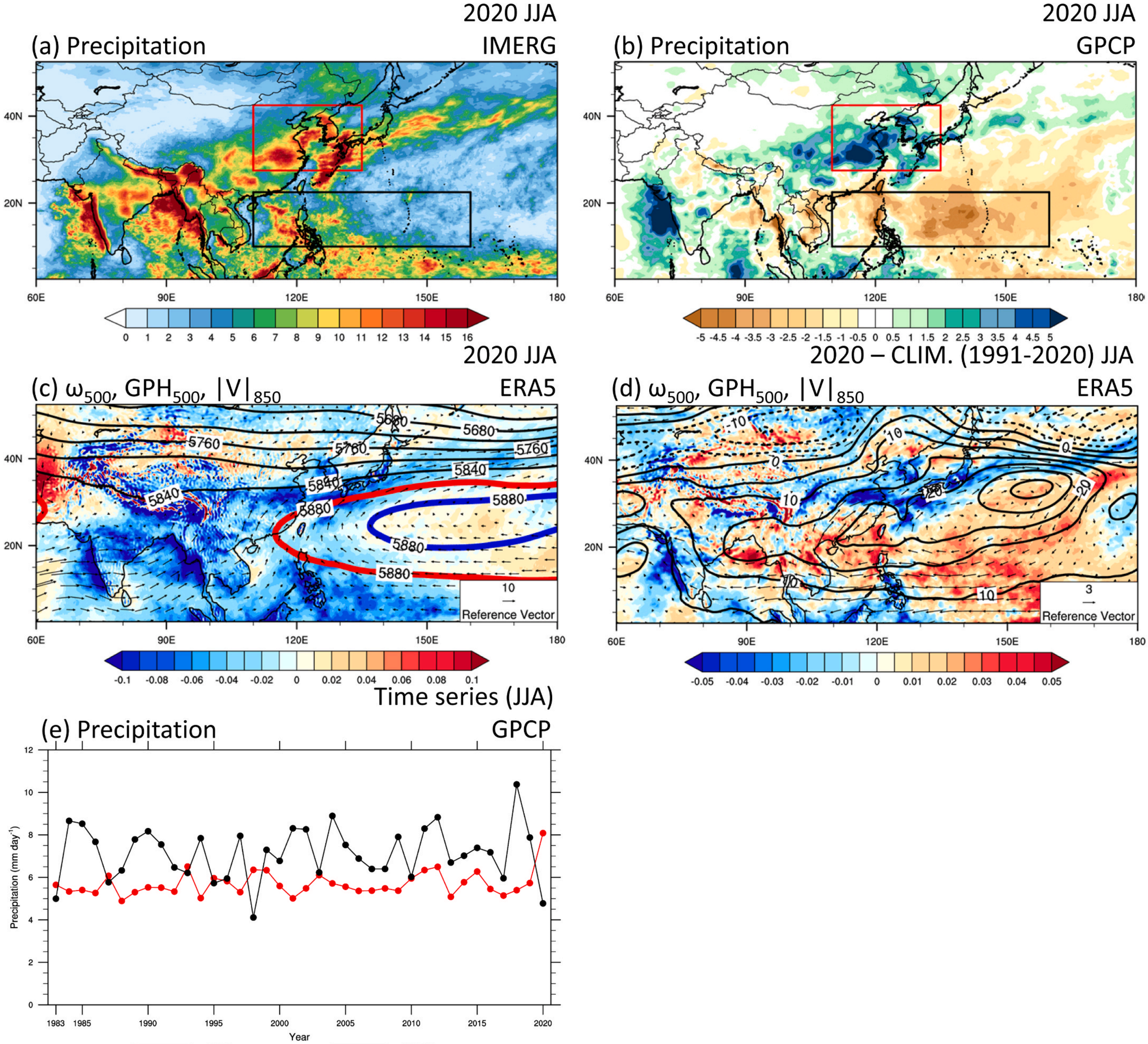

**Fig. 1.** (a) Average and (b) anomaly of daily mean precipitation (mm day$^{-1}$) during 2020 June–August (JJA) data adapted from monthly mean IMERG and GPCP. The red and black boxes in Figure (a) and (b) indicate EASM and WNPSM. ERA5 (c) average and (d) anomalies of 500 hPa geopotential height (gpm; contour) and omega (Pa s$^{-1}$; shading), and 850 hPa wind (m s$^{-1}$; vector) during 2020 JJA. The thick blue and red contour indicate 5880 gpm of JJA mean in climatology (1991–2020) and 2020. (e) Time series of GPCP daily mean precipitation (mm day$^{-1}$) at EASM (red) and WNPSM (black). (For interpretation of the references to colour in this figure legend, the reader is referred to the Web version of this article.)

The changes in SST can affect the variability of the EASM. Various studies are underway on how the increasing SST trend associated with global warming affects the EASM climate using both observations and climate model simulations (Fan et al., 2013; Ham et al., 2021; Li et al. 2010a, 2010b; Zhu et al., 2012). Note that not only global warming but also natural long-term variabilities such as the Pacific Decadal Oscillation (PDO) and Interdecadal Pacific Oscillation (IPO) may have contributed to the recent SST warming pattern (Gu et al., 2016; Meehl et al., 2016). Trenberth et al. (2003) stated the moisture-holding capacity of air changes to 7 % $K^{-1}$ due to the Clausius-Clapeyron relationship, and precipitation can increase due to this thermodynamic effect. In addition to the thermodynamic effects of SST warming, dynamic effects such as changes in low-level moisture convergence due to transport can affect local precipitation (Cha et al., 2011; Trenberth et al., 2003; Trenberth 2011). Despite the apparent increase in SST in the western North Pacific (WNP) (Fig. 2), there have been a few studies on how it affected high-impact weather and climate events, such as the extreme EASM in 2020. Therefore, in this study, we investigate the impact of the recent SST warming pattern on the large-scale circulation associated with the extreme EASM in 2020 and its thermodynamic and dynamic effects, using a regional climate model, which has the added value of simulating extreme weather and climate events more accurately with high-resolution and detailed topography (Cha et al., 2016).

## 2. Data and experiments

### 2.1. Data and method

This study used the monthly mean precipitation data of the Integrated Multi-satellitE Retrievals for GPM (IMERG) V06 0.1° × 0.1° to investigate the precipitation of the 2020 summer (Huffman et al., 2019). As IMERG data has been available since 2000, we used monthly mean precipitation analysis of the Global Precipitation Climatology Project (GPCP) Version 3.2 on a 0.5° × 0.5° grid from 1983 to 2020 for the climatological analysis of precipitation (Huffman et al., 2022). To evaluate the model performance, we used the European Centre for Medium-Range Weather Forecasts (ECMWF) Reanalysis Version 5 (ERA5) on a 0.25° × 0.25° horizontal grid with monthly and 6-h temporal resolutions from 1979 to 2020 (Hersbach et al., 2020). The SST analysis is conducted by using ERA5, National Oceanic and Atmospheric Administration (NOAA) Extended Reconstructed Sea Surface Temperature (ERSST) Version 5 with 2.0°×2.0° resolution (Huang et al., 2017Huang et al 2017), and Hadley Centre Sea Ice and Sea Surface Temperature data set (HadISST) with 1.0°×1.0° resolution monthly data for 42 years (1979-2020) (Rayner et al., 2003).

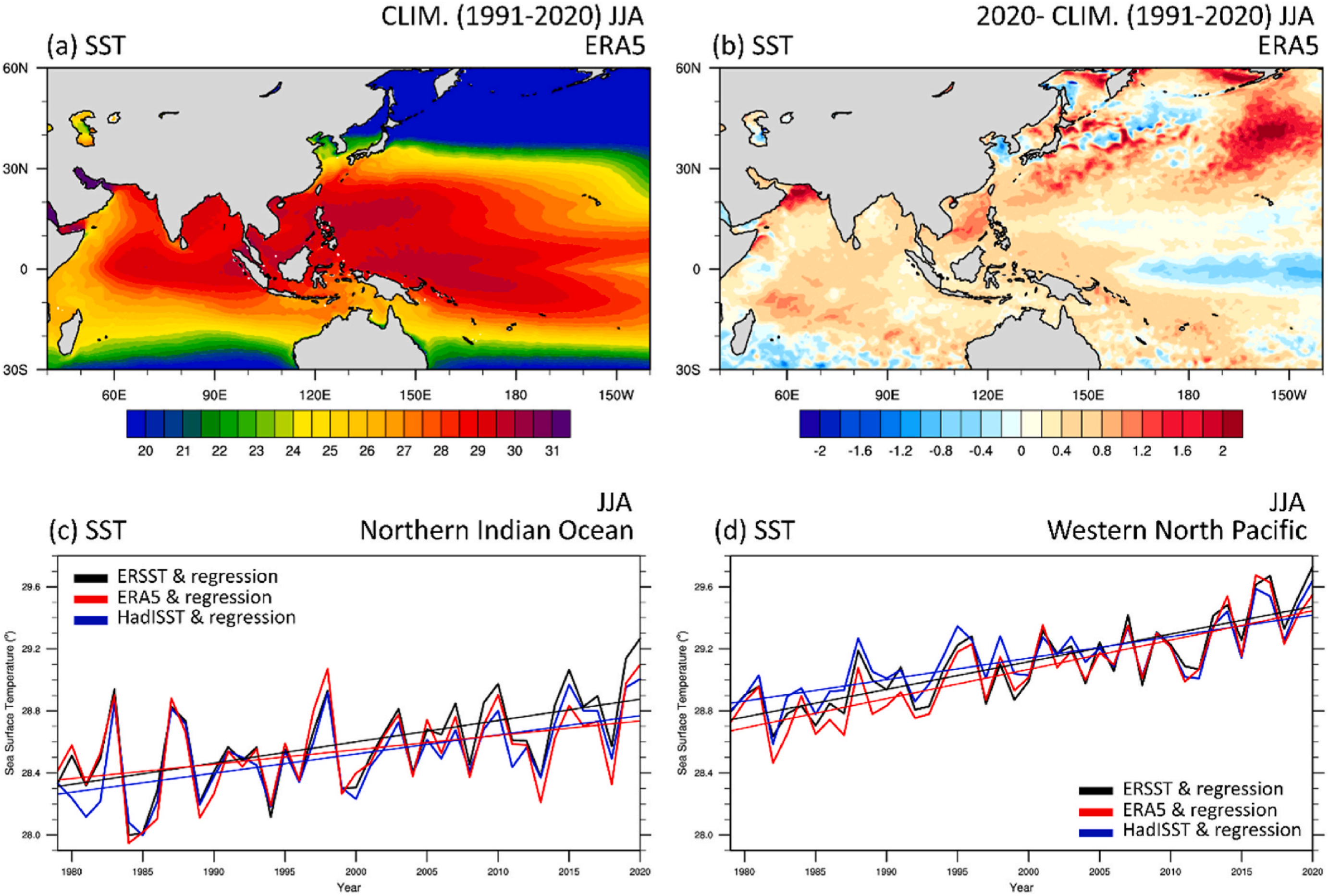

**Fig. 2.** (a) The climatological mean of SST and (b) 2020 anomaly with ERA5 data. The time series of mean sea surface temperature of (c) the northern Indian Ocean and (d) the western North Pacific calculated by ERSST (black), ERA5 (red), and HadISST (blue) data and their linear regression trend during JJA. Mean sea surface temperatures in the northern Indian Ocean and the western North Pacific are calculated for the ocean regions of 0°N–25°N, 50°E–110°E, and 0°N–25°N, 120°E–180°E, respectively. (For interpretation of the references to colour in this figure legend, the reader is referred to the Web version of this article.)

It is well known that the EASM is associated with the WNPSM as a dipole mode (Cha and Lee 2009; Kwon et al., 2005; Wang and LinHo 2002). Therefore, we quantitatively analyzed precipitation related to the EASM and the WNPSM. Based on previous studies (Ha et al., 2018; Li et al., 2014; Murakami and Matsumoto 1994; Oh et al., 2018; Wang et al., 2008; Yihui and Chan 2005) and the spatial distribution of daily mean precipitation in the climatological summer (1991–2020 June, July, and August; JJA), we defined the two monsoon regions for the quantitative analysis of precipitation in both monsoons: the EASM (27.5 °N–42.5 °N, 110 °E–135 °E) and the WNPSM (10 °N–22.5 °N, 110 °E−160 °E) (Fig. 1a).

To investigate the characteristics of EASM, Zeng et al. (2007) examined the local Walker and Hadley circulations to comprehend the circulation anomalies. In this study, the vertical and horizontal winds will be referred to as the local Walker and Hadley circulation, respectively. They are defined to examine the zonal and meridional circulation changes, including the enhancement and reduction of convection. Note that these definitions might differ from those used by other authors (Schwendike et al., 2014; Wang 2002).

The model performance was verified using pattern correlation coefficient, bias, and root mean square error (RMSE), and the long-term trends were analyzed using linear regression and Student's t-test.

### 2.2. Experiments

This study used the Weather Research and Forecast (WRF) Model version 4.1.2 for numerical modeling experiments (Skamarock et al., 2019). This study aimed to analyze the potential impact of the recent SST warming and whether it strengthens the EASM precipitation. Three experiments were particularly designed (Table 1). First, a control experiment (CTL) was conducted by prescribing the SST from the ERA5. The model was initialized from 00 UTC on May 1 to 00 UTC on May 3, 2020, at 12-h intervals to generate five ensemble members. Each ensemble member was integrated to August 30, 2020, and analyzed for JJA. Two sensitivity experiments are also conducted by modifying SST. The monthly SST trend was calculated at each grid point using the ERA5 SST data from 1979 to 2020. The resulting trend was subtracted from or added to the 2020 SST. They are referred to as cold SST experiment (CSST) and warm SST experiment (WSST; Figure S1). These two experiments were conducted to investigate the relationship between the recent SST warming and the EASM precipitation.

**Table 1**
Model Configurations.

| Model | Weather Research and Forecast (WRF) Model V4.1.2 |
|---|---|
| Control experiments (CTL) | Original ERA5 SST |
| Cool SST experiments (CSST) | Original ERA5 SST minus 22-year SST monthly trend |
| Warm SST experiments (WSST) | Original ERA5 SST plus 22-year SST monthly trend |
| Simulation period | 2020-05-01, 00 UTC to 2020-09-01, 00 UTC<br>2020-05-01, 12 UTC to 2020-09-01, 00 UTC<br>2020-05-02, 00 UTC to 2020-09-01, 00 UTC<br>2020-05-02, 12 UTC to 2020-09-01, 00 UTC<br>2020-05-03, 00 UTC to 2020-09-01, 00 UTC |
| Analysis period | 2020-05-31, 15 UTC to 2020-08-31, 15 UTC |
| Horizontal grids (resolution) | 1441 × 571 (12 km) |
| Vertical grids (model top) | 35 levels (50 hPa) |
| Initial/boundary condition | ECMWF Reanalysis V5 (ERA5) data six hourly, 0.25° horizontal resolution |
| Microphysics parameterization | WSM6 (Hong and Lim 2006) |
| Cumulus parameterization | KSAS (Kwon and Hong 2017) |
| Radiation-shortwave | RRTMG (Iacono et al., 2008) |
| Radiation-longwave | RRTMG (Iacono et al., 2008) |
| Planetary boundary layer | YSU (Hong et al., 2006) |
| Land surface model | Noah Land Surface Model (Tewari et al., 2004) |

To mitigate the effect of interannual and interdecadal variability of SST, we calculated the SST trend for 42 years (1979–2020) and subtracted or added the value for 22 years to investigate the effect of the recent SST warming pattern between the two years (i.e., 1998 and 2020) with the extremely strong EASM, as the record-breaking monsoon rainfall occurred in 1998 and 2020 (Lee et al., 2005). The differences in JJA mean SST in the WNP (0 °N–25 °N, 120 °E−180 °E) and northern Indian Ocean (NIO; 0 °N–25 °N, 50 °E−110 °E) regions between the CTL and CSST are about 0.42 K and 0.20 K, respectively. This indicates that the recent SST warming trend is stronger in the WNP region, increasing the SST gradient between the two basins (Fig. 2a–c and d).

## 3. Results

The CTL simulates the precipitation reasonably during the summer of 2020 compared to IMERG precipitation data (Figs. 1a and 3a). The CTL qualitatively reproduces the spatial patterns of the EASM, WNPSM, and Indian Summer Monsoon similar to the observation. The average daily precipitation in the EASM and WNPSM regions are 8.72 and 5.27 mm day$^{-1}$ in the IMERG monthly data, simulated as 6.65 and 6.16 mm day$^{-1}$ (Table 2). Through an optimization test for the cumulus parameterization scheme (CPS), it was confirmed that the KSAS scheme tends to somewhat underestimate the EASM precipitation but simulates the EASM, tropical, and other monsoon precipitation patterns better than other CPS options such as (e.g., Kain-Fritsch, multi-scale Kain-Fritsch, Tiedtke; not shown). The pattern correlations between the GPCP monthly data and the CTL precipitation of the entire domain, EASM, and WNPSM region are 0.79, 0.67, and 0.63 respectively. In the CTL, therefore, EASM precipitation is underestimated, but the pattern correlation is high enough. It indicates that the characteristics of EASM precipitation affected by various drivers (e.g., ENSO and IPOC) are properly reproduced by the model. Unlike EASM precipitation, WNPSM precipitation is overestimated. Despite these biases in precipitation simulation, the CTL shows that the WRF model has the ability to capture the overall distribution of EASM precipitation in 2020.

The difference between the CTL and CSST reveals the effect of the SST warming trend on the precipitation change. Fig. 3b shows that precipitation in the CTL was larger than that in the CSST by 1.17 mm day$^{-1}$ (19.0 %) in the WNPSM region but smaller by −0.32 mm day$^{-1}$ (−4.8 %) in the EASM region (Table 2). The increase in the WNPSM precipitation is due to a thermodynamic effect caused by warm SST, which is related to the moisture-holding capacity of the atmosphere around 7 % K$^{-1}$ (i.e., the Clausius-Clapeyron relationship). The JJA mean SST difference between the CTL and CSST for the EASM and WNPSM regions is about 0.70 K and 0.39 K, respectively. The precipitation change in the EASM region is smaller than that in the WNPSM region. This result is consistent with previous studies documenting that tropical precipitation is more sensitive to SST change than mid-latitude precipitation (e.g., O'Gorman 2012). Note that precipitation increases in the WNPSM region but decreases in the EASM region with the recent warm SST pattern. The precipitation difference between CTL and CSST (Fig. 3b) shows the exact opposite pattern to the precipitation anomaly in 2020 summer (Fig. 1b). It may suggest that recent SST warming may not contribute to precipitation increase in the EASM region. It further suggests that SST change has not only the thermodynamic effect that increases the precipitation but also the dynamic effects that reduce the precipitation. To investigate the mechanisms for the heavy precipitation during the summer of 2020, we analyzed the composite of seasonal mean synoptic fields (i.e., 500 hPa geopotential height and omega, and the 850 hPa wind). To identify areas where the frontal line and extreme precipitation occur, we use the 5880 gpm line of geopotential height as a criterion for the edge of the WNPSH similar to previous studies (He et al., 2015; Yoon et al., 2018). In 2020 summer, the WNPSH expanded northwestward significantly compared to the climatological average, and the 850 hPa zonal wind converged near the Philippines which

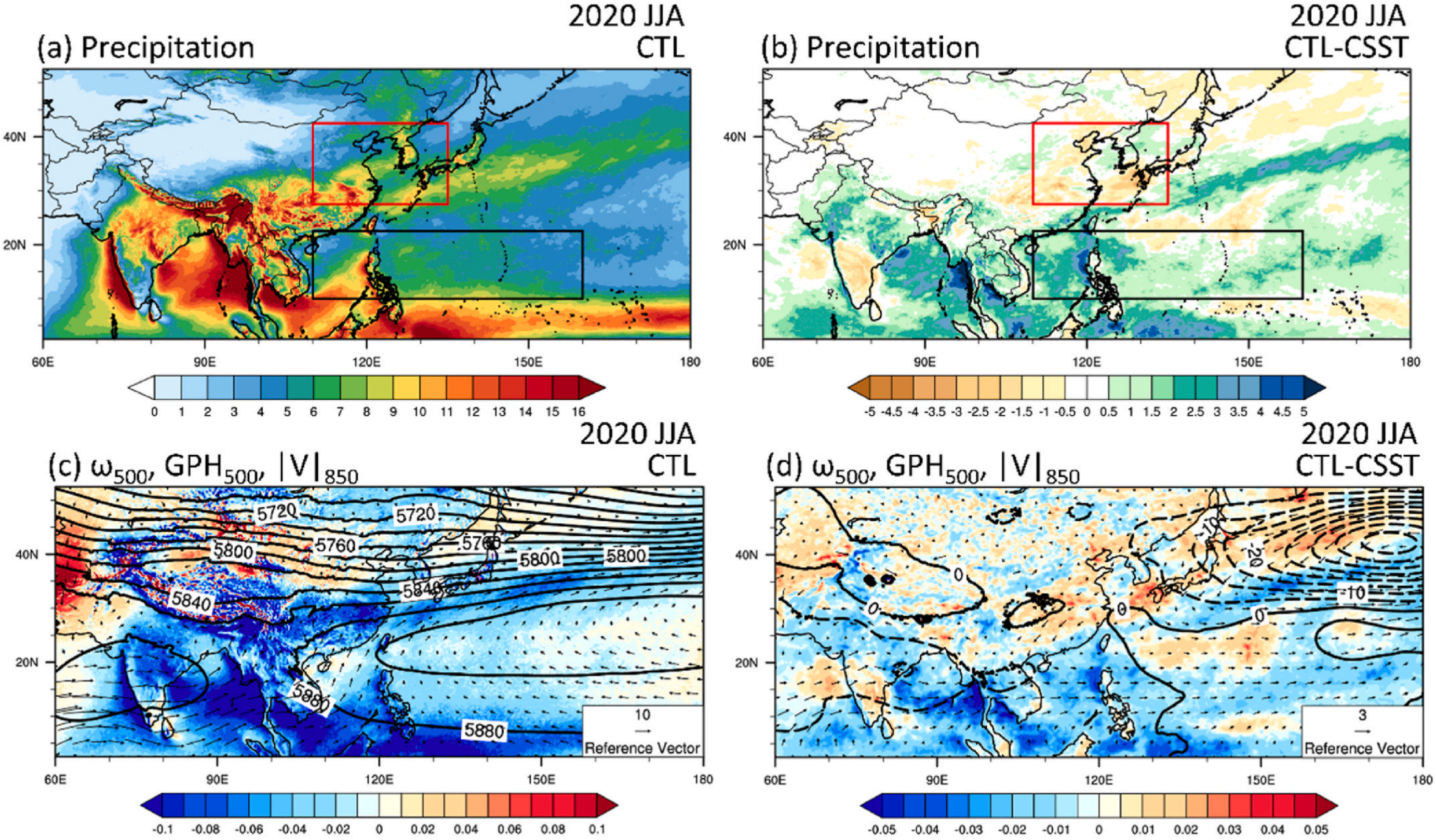

**Fig. 3.** Daily mean precipitation (mm day$^{-1}$) of (a) CTL ensemble means and (b) difference between CTL and CSST ensembles during 2020 JJA. 500 hPa geopotential height (gpm; contour) and omega (Pa s$^{-1}$; shading), and 850 hPa wind (m s$^{-1}$; vector) of (c) CTL ensemble means and (d) difference between CTL and CSST ensembles during 2020 JJA.

**Table 2**
Daily mean precipitation of IMERG and CTL ensemble, and differences of precipitation between CTL and CSST or WSST ensembles at EASM and WNPSM.

| EXP | IMERG | | CTL | | CTL - CSST | | CTL - WSST | |
|---|---|---|---|---|---|---|---|---|
| region | EASM | WNPSM | EASM | WNPSM | EASM | WNPSM | EASM | WNPSM |
| Precipitation (rate of change) | 8.72 | 5.27 | 6.65 | 6.16 | −0.32 (−4.8%) | 1.17 (19.0%) | 0.40 (5.9%) | −2.99 (−48.5%) |

unit: mm day$^{-1}$.

supplied moisture to East Asia along the edge of the WNPSH where the strong ascending motion occurred (Fig. 1c). The anomalous patterns of northwestward expansion of the WNPSH, strong convergence of low-level winds at the Philippine Sea (0°N-20°N, 120°E-130°E), and strong convection in East Asia are also properly reproduced in the CTL (Fig. 3c). Fig. 3d shows the difference in synoptic fields between the CTL and CSST, where the edge of the WNPSH in the CTL does not expand more northwestward than in the CSST, indicating the shrinkage of the WNPSH in the CTL. In the CTL and CSST differences, a negative omega difference is dominant in the tropics, while a positive one prevails in the mid-latitude; this implies that the recent warm SST pattern enhances tropical convection, especially on the Indochina Peninsula and the Philippines coast, but suppresses extratropical convection in the EASM region. Also, a cyclonic circulation difference at the low-level appeared over southern China. Therefore, the warm SST weakens the moisture transport to the EASM region compared to the CSST. This pattern is contrary to the 2020 anomaly (see Fig. 1d).

This result suggests that the increase in SST played a role in decreasing the monsoon precipitation by suppressing the anomalous synoptic pattern of the extreme EASM precipitation in 2020. The atmospheric circulation pattern change can be explained as follows. First, in summer, the SST in the WNP is climatologically higher than the SST in the northern Indian Ocean. In addition, the observed SST increase is greater or similar in the WNP than in the IO. It increases the SST gradient between the two regions (Fig. 2), resulting in stronger low-level westerly winds from IO and WNP in the CTL than the CSST. The warm SST and increased low-level westerly winds significantly enhance convection in the Bay of Bengal (BoB; 0°N-20°N, 80°E-100°E) and the South China Sea (SCS; 0°N-20°N, 105°E-120°E) and suppress the westward expansion of the WNPSH.

SST sensitivity experiments suggest that convective activity and wind field in the equatorial region and mid-latitude can change, affecting large-scale circulations. Therefore, we first analyzed the effect of increased SST on the local Walker circulation (Fig. 4a–d). Although the CTL simulates the tropic convection slightly stronger than the ERA5, the spatial pattern of the local Walker circulation is properly reproduced (Fig. 4a and b). The CTL simulates intensified convection in the BoB and SCS compared to the CSST (Fig. 4d). Also, stronger westerly (easterly) at the lower (upper) troposphere exhibits in the CTL than in the CSST. Therefore, the CTL enhances the local Walker circulation between the IO and the WNP compared to the CSST. In addition, the WNPSH, which can suppress the convection in the WNPSM region, is contracted, so the low-latitude precipitation in the CTL with higher SST is greater than that in the CSST (Fig. 3b and d). The difference in the direction of low-latitude zonal and vertical wind between the CTL and CSST (Fig. 4d) is opposite to that of the 2020 anomaly (Fig. 4c). These results imply that the increase in SST could enhance the local Walker circulation by intensifying convection in the tropics.

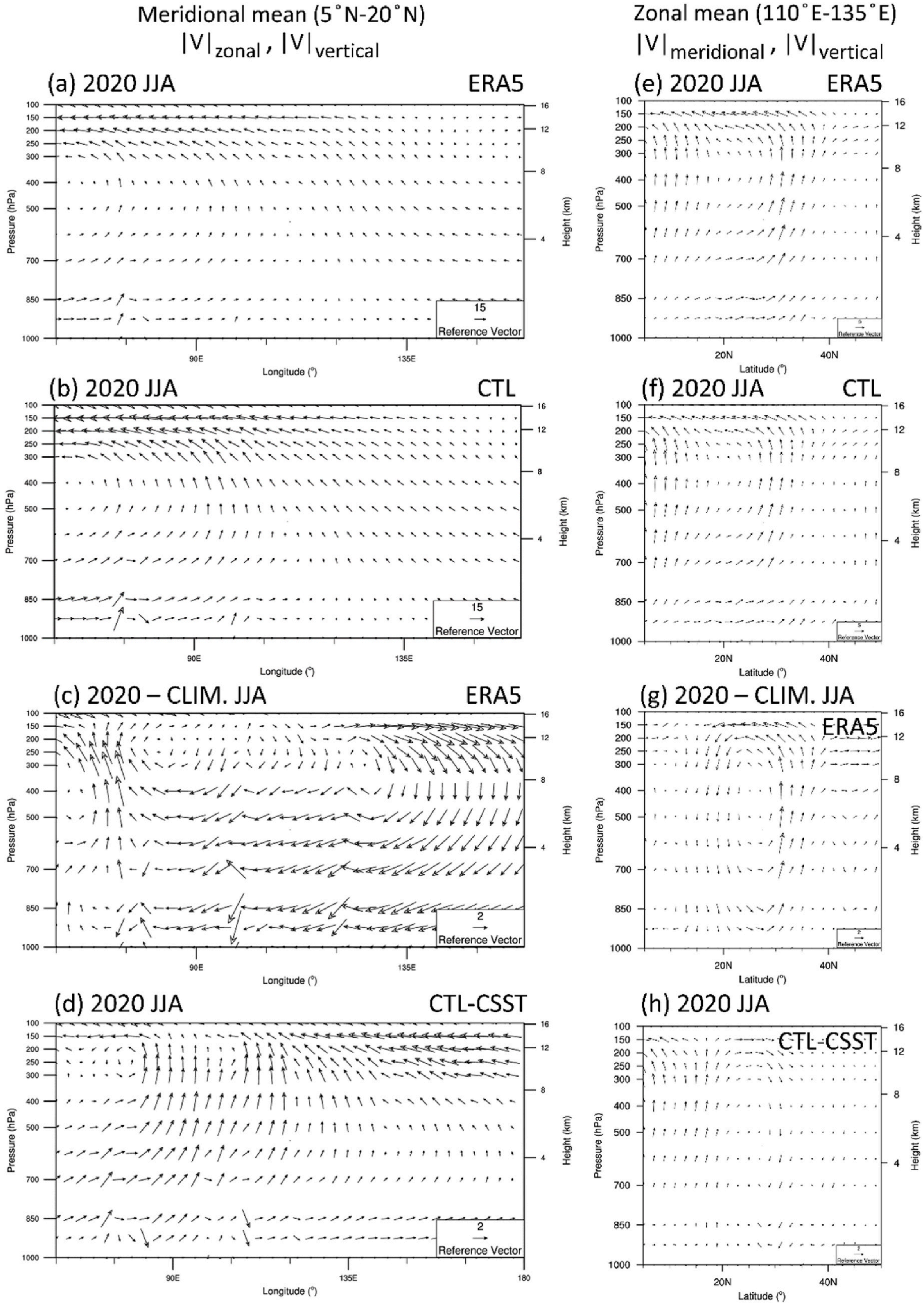

**Fig. 4.** The meridional mean (5N–20N) zonal wind (m $s^{-1}$) and vertical velocity ($10^2$ m $s^{-1}$) of (a) ERA5, and (b) ensemble mean of CTL runs. (c) ERA5 anomaly in 2020 JJA and (d) differences between CTL and CSST ensembles mean of the meridional mean (5N–20N) zonal wind (m $s^{-1}$) and vertical velocity ($10^2$ m $s^{-1}$). The zonal mean (110E-135E) meridional wind (m $s^{-1}$) and vertical velocity ($10^2$ m $s^{-1}$) of (e) ERA5, and (f) ensemble mean of CTL runs. (g) ERA5 anomaly in 2020 JJA and (h) differences between CTL and CSST ensembles mean of the zonal mean (110E-135E) meridional wind (m $s^{-1}$) and vertical velocity ($10^2$ m $s^{-1}$). All analyses were conducted during 2020 JJA.

The effect of SST on the local Hadley circulation is analyzed (Fig. 4e–h). Compared to the ERA5 reanalysis field, the overall vector flow and intensity are well simulated in the CTL (Fig. 4e and f). In the CTL, convection in the tropics is increased while convection in mid-latitude is suppressed. Therefore, the clockwise circulation is intensified compared to those in the CSST (Fig. 4h). In contrast, the 2020 anomaly is a pattern with a strong counter-clockwise circulation with increasing convection in mid-latitude (Fig. 4g). The strengthening of meridional wind flow that circulates clockwise indicates the intensification of the local Hadley circulation. This result indicates that the increasing SST pattern played a role in suppressing the anomalous Hadley circulation in 2020, which was weaker than climatology. Previous studies have also suggested the mechanisms for the reinforced Walker circulation and the enhanced Hadley cell by changes in the SST gradient (Li et al., 2016; Zaplotnik et al., 2022). Zeng et al. (2007) conducted a comparison of interdecadal variations during strong and weak EASM periods. They suggested that the presence of a warm SST anomaly in the eastern tropical Pacific region produced an abnormal ascending motion, which in turn strengthened the Walker cell, generated an anti-Walker cell to the west (upward motion near 100°E), and strengthened the East Asia Hadley cell. This process ultimately resulted in a weakening of EASM precipitation.

In addition, a moist budget analysis is conducted to clarify the dynamic and thermodynamic contributions of the SST increase effect on extreme EASM precipitation. The moist budget equation of the atmosphere with horizontal diffusion is as follows (Schmitz and Mullen 1996):

$$E - P = \frac{\partial W}{\partial t} + \nabla \bullet Q \tag{1}$$

where E and P represent the evaporation and precipitation, W is the precipitable water, and Q is the vertically integrated water vapor flux:

$$Q = -\frac{1}{g}\int_{Psfc}^{Ptop} qV\,dp \tag{2}$$

where g represents the gravitational constant, $p_{top}$ and $p_{sfc}$ are the pressure at the top of the troposphere and surface pressure, and V and q represent horizontal wind and specific humidity, respectively. Furthermore, we calculated the moisture transported through each boundary $Q_b$ of the EASM region (Piao et al., 2018):

$$Q_b = \int_L Q \times n\,dl \tag{3}$$

where n is the normal vector into the EASM box, and L represents the length of each boundary.

The changes in precipitation, evaporation, and convergence are investigated in the sensitivity experiment and the sum of $Q_b$ is expressed as convergence. The CTL confirms that moisture is transported into the southern and western boundaries of the EASM region and that moisture outflow from the eastern and northern boundaries (Fig. 5a). In the CTL with warmer SST, water vapor convergence from the northern boundary increases by 0.62 mm day$^{-1}$ compared to the CSST, but decreases from the southern, eastern, and western boundaries by 0.84, 0.13, and 0.15 mm day$^{-1}$, respectively. Hence, the net convergence decreases by 0.49 mm day$^{-1}$ (Fig. 5b). In other words, the supply of water vapor into the EASM region due to dynamic components decreases in the CTL compared to the CSST. However, the CTL simulates more evaporation in East Asia by 0.17 mm day$^{-1}$ due to warmer SST compared to the CSST. In the summer of 2020, evaporation in the EASM region increased due to warmer SST. Still, a greater amount of convergence decreased, inducing a decrease in precipitation.

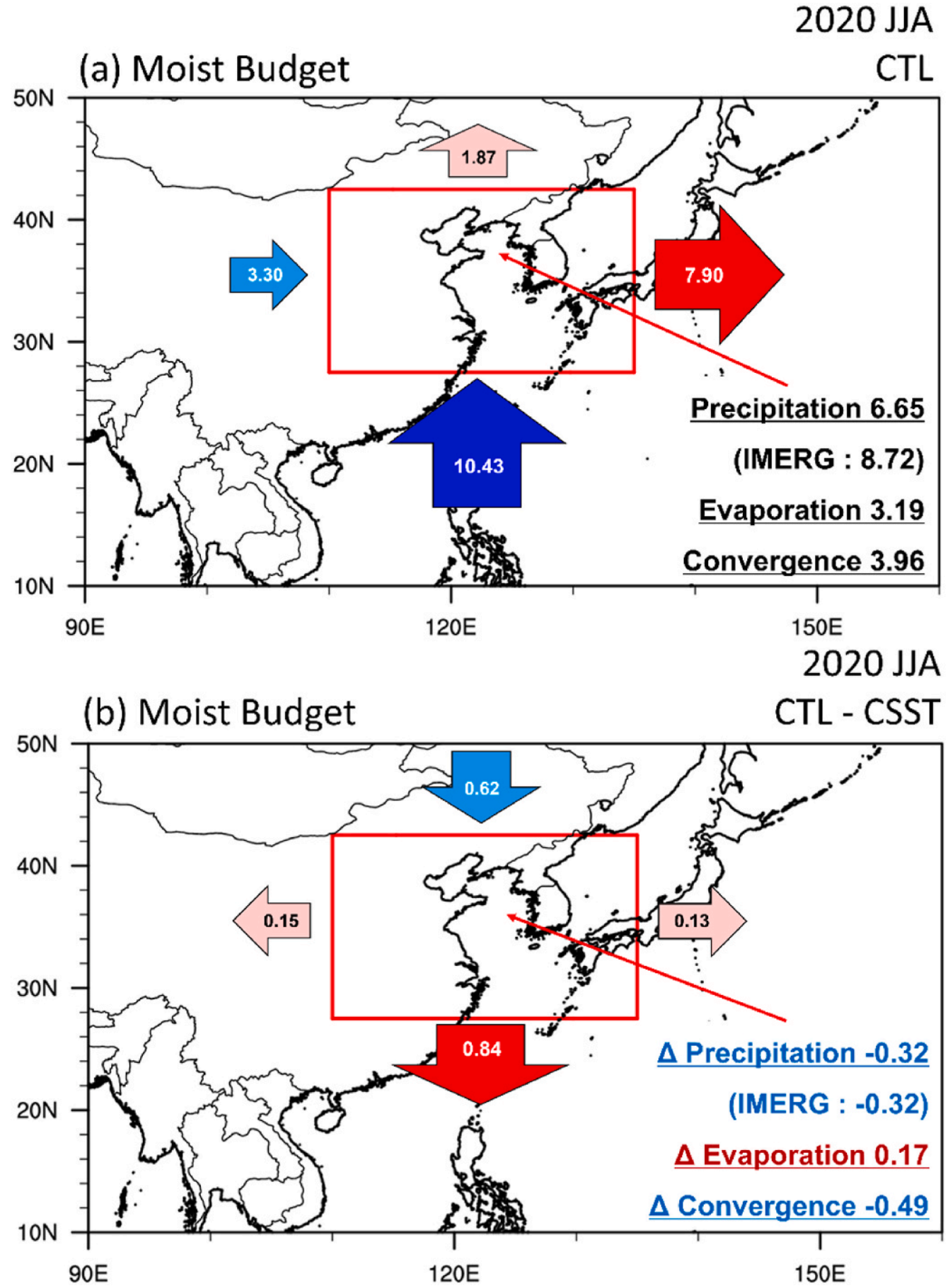

**Fig. 5.** The moist budget for EASM of (a) CTL and (b) differences between CTL and CSST through each boundary. The numbers in the blue arrows indicate the amount of water vapor (mm day$^{-1}$) that is convergence into the EASM region at each boundary, and red, vice versa. (For interpretation of the references to colour in this figure legend, the reader is referred to the Web version of this article.)

## 4. Summary and conclusion

This study investigated the dynamic and thermodynamic effects of recent SST warming on the extreme EASM precipitation in 2020. During the summer of 2020, the WNPSH expanded significantly to the northwestward, and sufficient water vapor was supplied along the edge of the WNPSH due to high SST, allowing the monsoon rain band and extratropical cyclone to record high precipitation in East Asia. However, the trend of increasing SST in summer was higher in the WNP than in the IO, which increased the SST gradient between the two regions. The CSST with a less advanced trend of increasing SST was compared with the CTL. The CTL using the WRF model quantitatively captured the precipitation characteristics of East Asia, the WNP, and the entire experimental domain. The effect of increasing SST represented by the difference between CTL and CSST, induced the increased SST gradient between the IO and WNP, which enhanced low-level westerlies between them and contracted the WNPSH (Fig. 6). Due to the changes in these synoptic fields, the convergence of the local Walker circulation increased in the SCS and Philippine Sea, while warmer SST strengthened convection in the tropics, resulting in more precipitation. Simultaneously, the convection in the low-latitude area of the local Hadley circulation was enhanced by the warmer SST, which strengthened the circulation and suppressed the convection in mid-latitude (i.e., East Asia). As a result, the average precipitation in the EASM region decreased in inverse proportion to the SST change due to the dynamic effect, which was greater than the thermodynamic effect. In contrast, the average precipitation in the WNPSM region increased because of the sensitive response to the thermodynamic effect of SST change. These effects showed opposite patterns to the 2020 anomaly that caused the extreme EASM precipitation. A WSST experiment with a more advanced SST trend was also conducted to verify the SST warming effect further. When comparing the differences between the WSST and CTL experiments, the differences in precipitation and synoptic patterns were similar to those between the CTL and CSST (see Figure S2-4), confirming that the dynamic effect of warmer SST on the extreme EASM in 2020 was greater than the thermodynamic effect.

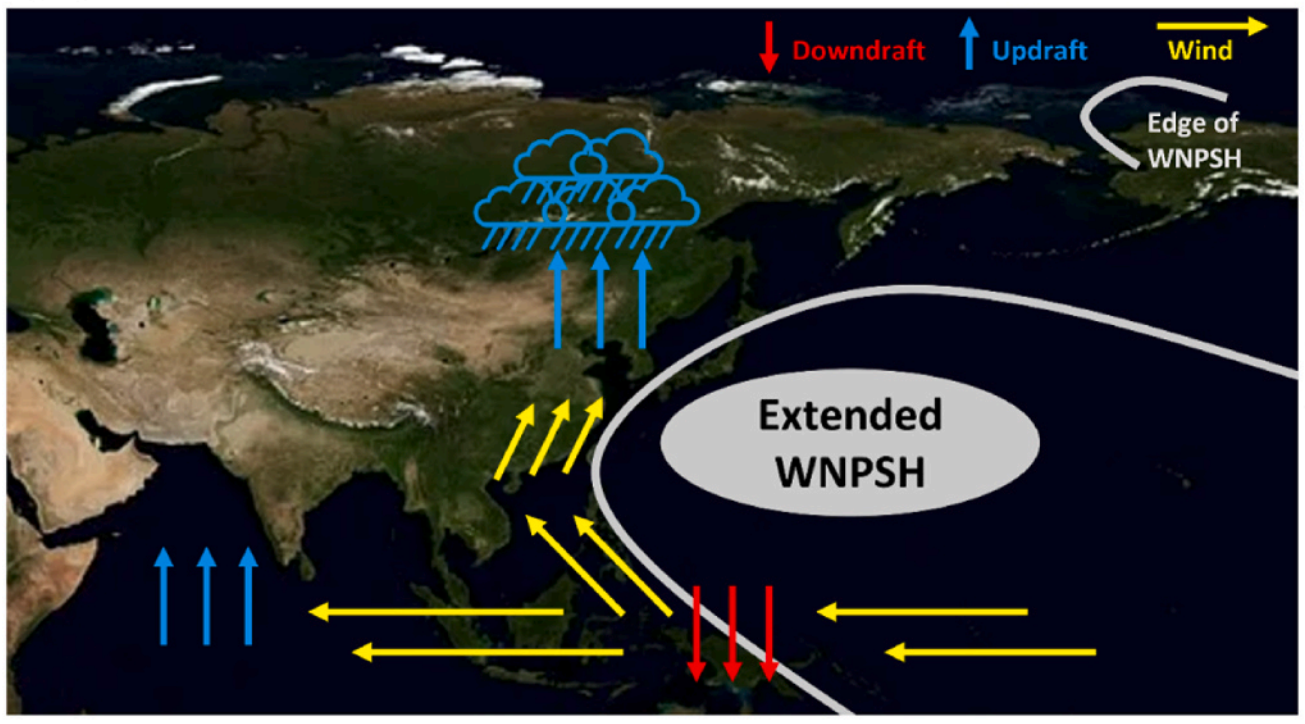

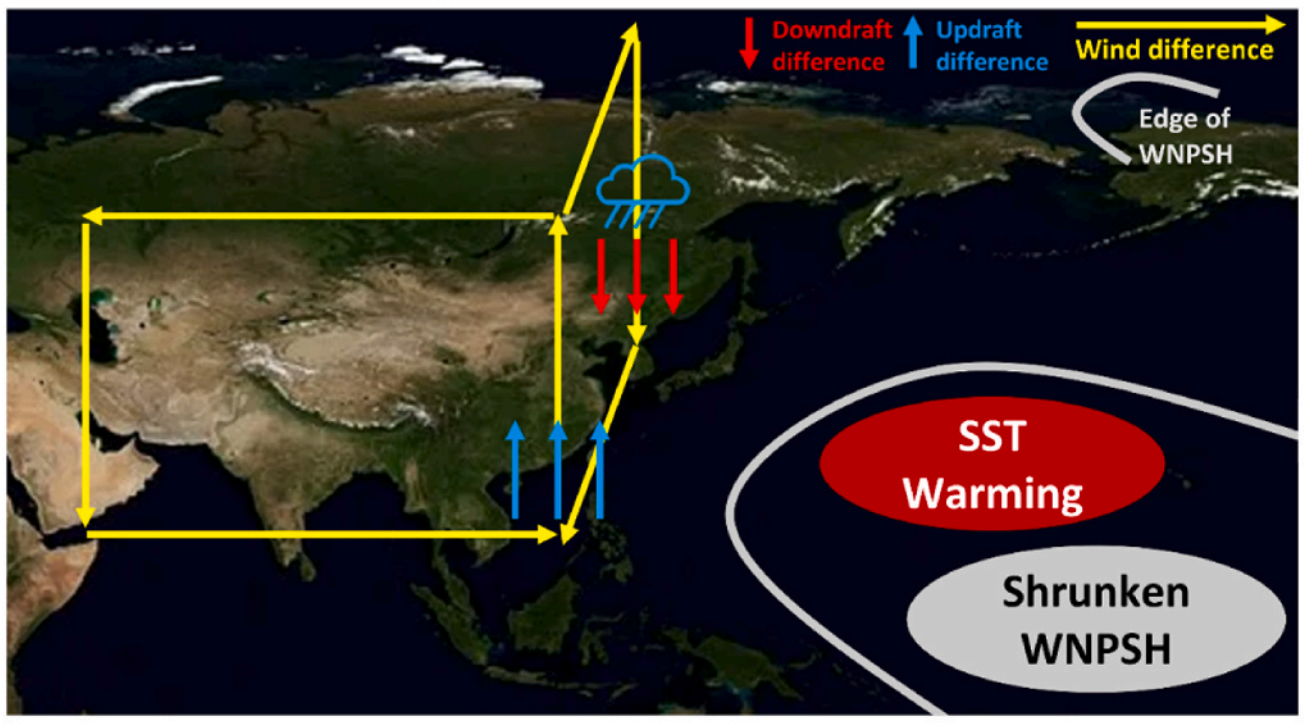

**Fig. 6.** (a) A schematic diagram of the synoptic pattern contributing to EASM precipitation in 2020, and (b) of the effect of the recent SST warming on the synoptic pattern in 2020 and the resulting change in EASM precipitation.

This study is meaningful because we provided the first quantification of the negative impact of recent SST warming on extreme EASM precipitation in 2020 using a high-resolution regional climate model. Although we focused on precipitation changes in the EASM region, our findings suggest that the impact of recent SST warming on precipitation changes in other regions also needs to be investigated. Furthermore, it is required to examine the uneven regional influences of SST warming on other high-impact weather events, such as tropical cyclones and heat waves.

Our analysis of the simulation ensemble indicates that the recent warming pattern of SST did not reinforce the extreme EASM in 2020, which resulted in record-breaking precipitation. Further experiments will be conducted to examine other possible mechanisms that caused the extreme EASM in 2020, including the Super IOD in 2019 and tropical North Atlantic warming, with reference to 1998, when strong external forcing resulted in past record-breaking EASM precipitation (Kim et al., 2022; Lee et al., 2005; Takaya et al., 2020). Because we used a regional climate model based on the limited area model to identify only the contribution of the recent SST warming to the physical mechanisms at the regional scale, it was difficult to examine the teleconnection effect of the recent SST warming pattern (e.g., the effects of the Arctic sea-ice loss and the North Atlantic warming) on the EASM in 2020. Therefore, a subsequent study will examine the teleconnection effect of SST warming. In addition, we will investigate the relationship between SST warming and the EASM for other years, such as weak or normal EASM years. Because a high-resolution model (e.g., convection-permitting model) can simulate extreme rainfall more accurately, it is necessary to conduct higher-resolution experiments to quantify the changes in extreme precipitation. Since we only identified the effect of the SST warming trend on EASM precipitation in 2020, we must investigate the impact of SST and atmospheric changes due to global warming by conducting the pseudo-global warming experiment (Kawase et al., 2009; Lee et al., 2023).

## CRediT authorship contribution statement

**Taeho Mun:** Writing – original draft, Visualization, Software, Resources, Methodology, Investigation, Formal analysis, Data curation. **Haerin Park:** Writing – review & editing, Visualization, Validation. **Dong-Hyun Cha:** Writing – original draft, Supervision, Project administration, Funding acquisition, Data curation, Conceptualization. **Chang-Keun Song:** Writing – review & editing, Supervision. **Seung-Ki Min:** Writing – review & editing, Supervision. **Seok-Woo Son:** Writing – review & editing, Supervision, Conceptualization.

## Declaration of competing interest

The authors declare that they have no known competing financial interests or personal relationships that could have appeared to influence the work reported in this paper.

## Data availability

Data will be made available on request.

The GPCP monthly precipitation data were obtained freely from https://measures.gesdisc.eosdis.nasa.gov/data/GPCP/GPCPMON.3.2/. The ERA5 data were downloaded from the Copernicus Climate Change Service Climate Data Store (CDS), https://cds.climate.copernicus.eu/cdsapp#!/home. The IMERG daily precipitation data are available at https://gpm1.gesdisc.eosdis.nasa.gov/data/GPM_L3/GPM_3IMERGDF.06/. The ERSST SST data are accessible at https://www.ncei.noaa.gov/pub/data/cmb/ersst/v5/netcdf/. The HadISST 406 data were obtained from https://www.metoffice.gov.uk/hadobs/hadisst/ and are © British 407 Crown Copyright, Met Office, 2022, provided under a Non-Commercial Government Licence http://www.nationalarchives.gov.uk/doc/non-commercial-government-licence/version/2/.

## Acknowledgments

This work was supported by the Korea Environment Industry & Technology Institute (KEITI) through the "Climate Change R&D Project for New Climate Regime.", funded by the Korea Ministry of Environment (MOE) (1485018907). We thank the supercomputing resources of the UNIST Supercomputing Center.

## Appendix A. Supplementary data

Supplementary data to this article can be found online at https://doi.org/10.1016/j.wace.2024.100682.

## References

Cai, Y., Chen, Z., Du, Y., 2022. The role of Indian Ocean warming on extreme rainfall in central China during early summer 2020: without significant El Niño influence. Clim. Dynam. 59, 951–960.

Cha, D.-H., Jin, C.-S., Lee, D.-K., 2011. Impact of local sea surface temperature anomaly over the western North Pacific on extreme East Asian summer monsoon. Clim. Dynam. 37, 1691–1705.

Cha, D.-H., Lee, D.-K., Hong, S.-Y., 2008. Impact of boundary layer processes on seasonal simulation of the East Asian summer monsoon using a regional climate model. Meteorol. Atmos. Phys. 100, 53–72.

Cha, D.-H., Jin, C.-S., Moon, J.-H., Lee, D.-K., 2016. Improvement of regional climate simulation of East Asian summer monsoon by coupled air–sea interaction and large-scale nudging. Int. J. Climatol. 36, 334–345.

Cha, D.-H., Lee, D.-K., 2009. Reduction of systematic errors in regional climate simulations of the summer monsoon over East Asia and the western North Pacific by applying the spectral nudging technique. J. Geophys. Res. Atmos. 114.

Chen, T.-J.G., Chang, C.-P., 1980. The structure and vorticity budget of an early summer monsoon trough (Mei-Yu) over southeastern China and Japan. Mon. Weather Rev. 108, 942–953.

Chen, X., Wen, Z., Song, Y., Guo, Y., 2022. Causes of extreme 2020 Meiyu-Baiu rainfall: a study of combined effect of Indian Ocean and Arctic *climate dynamics*, 59, 3485–3501.

Fan, L., Shin, S.I., Liu, Q., Liu, Z., 2013. Relative importance of tropical SST anomalies in forcing East Asian summer monsoon circulation. Geophys. Res. Lett. 40, 2471–2477.

Gu, G., Adler, R.F., Huffman, G.J., 2016. Long-term changes/trends in surface temperature and precipitation during the satellite era (1979–2012). Clim. Dynam. 46, 1091, 105.

Guo, Y., Zhang, R., Wen, Z., Li, J., Zhang, C., Zhou, Z., 2021. Understanding the role of SST anomaly in extreme rainfall of 2020 Meiyu season from an interdecadal perspective. Sci. China Earth Sci. 64, 1619–1632.

Ha, K.-J., Seo, Y.-W., Lee, J.-Y., Kripalani, R., Yun, K.-S., 2018. Linkages between the South and East Asian summer monsoons: a review and revisit. Clim. Dynam. 51, 4207–4227.

Ham, Y.-G., Kim, J.-G., Lee, J.-G., Li, T., Lee, M.-I., Son, S.-W., Hyun, Y.-K., 2021. The origin of systematic forecast errors of extreme 2020 East Asian Summer Monsoon rainfall in GloSea5. Geophys. Res. Lett. 48, e2021GL094179.

He, C., Zhou, T., Lin, A., Wu, B., Gu, D., Li, C., Zheng, B., 2015. Enhanced or weakened western North Pacific subtropical high under global warming? Sci. Rep. 5, 1–7.

He, C., Zhou, W., 2020. Different Enhancement of the East Asian summer monsoon under global warming and interglacial epochs simulated by CMIP6 models: role of the subtropical high. J. Clim. 33, 9721–9733.

Hersbach, H., Bell, B., Berrisford, P., Hirahara, S., Horányi, A., Muñoz-Sabater, J., Nicolas, J., Peubey, C., Radu, R., Schepers, D., 2020. The ERA5 global reanalysis. Q. J. R. Meteorol. Soc. 146, 1999–2049.

Hong, S.-Y., Lim, J.-O.J., 2006. The WRF single-moment 6-class microphysics scheme (WSM6). Asia-Pacific Journal of Atmospheric Sciences 42, 129–151.

Hong, S.-Y., Noh, Y., Dudhia, J., 2006. A new vertical diffusion package with an explicit treatment of entrainment processes. Mon. Weather Rev. 134, 2318–2341.

Huang, B., Thorne, P.W., Banzon, V.F., Boyer, T., Chepurin, G., Lawrimore, J.H., Menne, M.J., Smith, T.M., Vose, R.S., Zhang, H.-M., 2017. NOAA extended reconstructed sea surface temperature (ERSST), version 5. NOAA National Centers for Environmental Information 30, 8179–8205.

Huang, R., Zhou, L., Chen, W., 2003. The progresses of recent studies on the variabilities of the East Asian monsoon and their causes. Adv. Atmos. Sci. 20, 55–69.

Huffman, G.J., Behrangi, A., Bolvin, D.T., Nelkin, E.J., 2022. GPCP Version 3.2 Satellite-Gauge (SG) Combined Precipitation Data Set *Greenbelt, Maryland, USA*. Goddard Earth Sciences Data and Information Services Center (GES DISC).

Huffman, G.J., Stocker, E.F., Bolvin, D.T., Nelkin, E.J., Tan, J., 2019. GPM IMERG Final Precipitation L3 1 Day 0.1 Degree X 0.1 Degree V06 *Goddard Earth Sciences Data and Information Services Center (GES DISC)*.

Iacono, M.J., Delamere, J.S., Mlawer, E.J., Shephard, M.W., Clough, S.A., Collins, W.D., 2008. Radiative forcing by long-lived greenhouse gases: calculations with the AER radiative transfer models. J. Geophys. Res. Atmos. 113.

Kawase, H., Yoshikane, T., Hara, M., Kimura, F., Yasunari, T., Ailikun, B., Ueda, H., Inoue, T., 2009. Intermodel variability of future changes in the Baiu rainband estimated by the pseudo global warming downscaling method. Journal of geophysical research: Atmospheres 114.

Kim, S., Park, J.-H., Kug, J.-S., 2022. Tropical origins of the record-breaking 2020 summer rainfall extremes in East Asia. Sci. Rep. 12, 5366.

Kwon, M., Jhun, J.-G., Wang, B., An, S.-I., Kug, J.-S., 2005. Decadal change in relationship between east Asian and WNP summer monsoons. Geophys. Res. Lett. 32.

KMA, 2021. Abnormal climate report 2020 (in Korean). Korea Meteorological Administration 212, 30–37. http://www.climate.go.kr/home/bbs/view.php?code=93.

Kwon, Y.-C., Hong, S.-Y., 2017. A mass-flux cumulus parameterization scheme across gray-zone resolutions. Mon. Weather Rev. 145, 583–598.

Lau, K., Kim, K., Yang, S., 2000. Dynamical and boundary forcing characteristics of regional components of the Asian summer monsoon. J. Clim. 13, 2461–2482.

Lee, D.-K., Cha, D.-H., Choi, S.-J., 2005. A sensitivity study of regional climate simulation to convective parameterization schemes for the 1998 East Asian summer monsoon. TAO: Terr. Atmos. Ocean Sci. 16, 989.

Lee, M., Min, S.-K., Cha, D.-H., 2023. Convection-permitting simulations reveal expanded rainfall extremes of tropical cyclones affecting South Korea due to anthropogenic warming. npj Climate and Atmospheric Science 6, 176.

Li, H., Dai, A., Zhou, T., Lu, J., 2010a. Responses of East Asian summer monsoon to historical SST and atmospheric forcing during 1950–2000. Clim. Dynam. 34, 501–514.

Li, J., Wu, Z., Jiang, Z., He, J., 2010b. Can global warming strengthen the East Asian summer monsoon? J. Clim. 23, 6696–6705.

Li, X., Xie, S.-P., Gille, S.T., Yoo, C., 2016. Atlantic-induced pan-tropical climate change over the past three decades. Nat. Clim. Change 6, 275–279.

Li, Y., Wang, N a, Zhou, X., Zhang, C., Wang, Y., 2014. Synchronous or asynchronous Holocene Indian and East Asian summer monsoon evolution: a synthesis on Holocene Asian summer monsoon simulations, records and modern monsoon indices. Global Planet. Change 116, 30–40.

Liu, B., Yan, Y., Zhu, C., Ma, S., Li, J., 2020. Record-breaking Meiyu rainfall around the Yangtze River in 2020 regulated by the subseasonal phase transition of the North Atlantic Oscillation. Geophys. Res. Lett. 47, e2020GL090342.

Meehl, G.A., Hu, A., Santer, B.D., Xie, S.-P., 2016. Contribution of the Interdecadal Pacific Oscillation to twentieth-century global surface temperature trends. Nat. Clim. Change 6, 1005–1008.

Murakami, T., Matsumoto, J., 1994. Summer monsoon over the Asian continent and western North Pacific. Journal of the Meteorological Society of Japan. Ser. II 72, 719–745.

O'Gorman, P.A., 2012. Sensitivity of tropical precipitation extremes to climate change. Nat. Geosci. 5, 697–700.

Oh, H., Ha, K.J., Timmermann, A., 2018. Disentangling impacts of dynamic and thermodynamic components on late summer rainfall anomalies in East Asia. J. Geophys. Res. Atmos. 123, 8623–8633.

Pan, X., Li, T., Sun, Y., Zhu, Z., 2021. Cause of extreme heavy and persistent rainfall over Yangtze River in summer 2020. Adv. Atmos. Sci. 38, 1994–2009.

Park, C., Son, S.-W., Kim, H., Ham, Y.-G., Kim, J., Cha, D.-H., Chang, E.-C., Lee, G., Kug, J.-S., Lee, W.-S., 2021. Record-breaking summer rainfall in South Korea in 2020: synoptic characteristics and the role of large-scale circulations. Mon. Weather Rev. 149, 3085–3100.

Piao, J., Chen, W., Zhang, Q., Hu, P., 2018. Comparison of moisture transport between Siberia and northeast Asia on annual and interannual time scales. J. Clim. 31, 7645–7660.

Qian, C., Ye, Y., Zhang, W., Zhou, T., 2022. Heavy rainfall event in mid-August 2020 in southwestern China: contribution of anthropogenic forcings and atmospheric circulation *Bull. Am. Meteorol. Soc*, 103, S111–S117.

Rayner, N., Parker, D.E., Horton, E., Folland, C.K., Alexander, L.V., Rowell, D., Kent, E.C., Kaplan, A., 2003. Global analyses of sea surface temperature, sea ice, and night marine air temperature since the late nineteenth century. J. Geophys. Res. Atmos. 108.

Schmitz, J.T., Mullen, S.L., 1996. Water vapor transport associated with the summertime North American monsoon as depicted by ECMWF analyses. J. Clim. 9, 1621–1634.

Schwendike, J., Govekar, P., Reeder, M.J., Wardle, R., Berry, G.J., Jakob, C., 2014. Local partitioning of the overturning circulation in the tropics and the connection to the Hadley and Walker circulations. J. Geophys. Res. Atmos. 119, 1322–1339.

Skamarock, W.C., Klemp, J.B., Dudhia, J., Gill, D.O., Liu, Z., Berner, J., Wang, W., Powers, J.G., Duda, M.G., Barker, D.M., 2019. A Description of the Advanced Research WRF Model Version 4 National Center for Atmospheric Research, p. 145. Boulder, CO, USA.

Takaya, Y., Ishikawa, I., Kobayashi, C., Endo, H., Ose, T., 2020. Enhanced Meiyu-Baiu rainfall in early summer 2020: Aftermath of the 2019 super IOD event. Geophys. Res. Lett. 47, e2020GL090671.

Tao, S.-Y., Chen, L.-X., 1987. A review of recent research on East Asian monsoon in China. Monsoon Meteorology. Oxford University Press, London.

Tewari, M., Chen, F., Wang, W., Dudhia, J., LeMone, M., Mitchell, K., Ek, M., Gayno, G., Wegiel, J., Cuenca, R., 2004. Implementation and verification of the unified NOAH land surface model in the WRF model. In: 20th Conference on Weather Analysis and Forecasting/16th Conference on Numerical Weather Prediction, pp. 2165–2170.

Trenberth, K.E., 2011. Changes in Precipitation with Climate Change *Climate Research*, vol. 47, pp. 123–138.

Trenberth, K.E., Dai, A., Rasmussen, R.M., Parsons, D.B., 2003. The changing character of precipitation. Bull. Am. Meteorol. Soc. 84, 1205–1218.

Wang, B., LinHo, 2002. Rainy season of the Asian–Pacific summer monsoon. J. Clim. 15, 386–398.

Wang, B., Wu, Z., Li, J., Liu, J., Chang, C.-P., Ding, Y., Wu, G., 2008. How to measure the strength of the East Asian summer monsoon. J. Clim. 21, 4449–4463.

Wang, C., 2002. Atmospheric circulation cells associated with the El Niño–southern Oscillation. J. Clim. 15, 399–419.

Wang, X., Guan, Z., Jin, D., Zhu, J., 2021. East Asian summer monsoon rainfall anomalies in 2020 and the role of northwest Pacific Anticyclone on the Intraseasonal-to-interannual Timescales. J. Geophys. Res. Atmos. 126, e2021JD034607.

Xie, S.-P., Hu, K., Hafner, J., Tokinaga, H., Du, Y., Huang, G., Sampe, T., 2009. Indian Ocean capacitor effect on Indo–western Pacific climate during the summer following El Niño. J. Clim. 22, 730–747.

Yihui, D., Chan, J.C., 2005. The East Asian summer monsoon: an overview. Meteorol. Atmos. Phys. 89, 117–142.

Yoon, D., Cha, D.-H., Lee, G., Park, C., Lee, M.-I., Min, K.-H., 2018. Impacts of synoptic and local factors on heat wave events over southeastern region of Korea in 2015. J. Geophys. Res. Atmos. 123 (12), 81, 12,96.

Zaplotnik, Ž., Pikovnik, M., Boljka, L., 2022. Recent Hadley circulation strengthening: a trend or multidecadal variability? J. Clim. 35, 4157–4176.

Zeng, G., Sun, Z., Wang, W.-C., Min, J., 2007. Interdecadal variability of the East Asian summer monsoon and associated atmospheric circulations. Adv. Atmos. Sci. 24, 915–926.

Zhang, R., Sumi, A., Kimoto, M., 1999. A diagnostic study of the impact of El Niño on the precipitation in China. Adv. Atmos. Sci. 16, 229–241.

Zhang, W., Huang, Z., Jiang, F., Stuecker, M.F., Chen, G., Jin, F.F., 2021. Exceptionally persistent Madden-Julian Oscillation activity contributes to the extreme 2020 East Asian summer monsoon rainfall. Geophys. Res. Lett. 48, e2020GL091588.

Zhou, T., Ren, L., Zhang, W., 2021. Anthropogenic influence on extreme Meiyu rainfall in 2020 and its future risk. Sci. China Earth Sci. 64, 1633–1644.

Zhu, C., Wang, B., Qian, W., Zhang, B., 2012. Recent weakening of northern East Asian summer monsoon: a possible response to global warming. Geophys. Res. Lett. 39.

Supporting Information for

# Did recent sea surface temperature warming reinforce the extreme East Asian Summer Monsoon precipitation in 2020?

Taeho Mun[1], Haerin Park[1], Dong-Hyun Cha[1*], Chang-Keun Song[1], Seung-Ki Min[2], and Seok-Woo Son[3]

[1] *Department of Civil, Urban, Earth and Environmental Engineering, Ulsan National Institute of Science and Technology, Ulsan, South Korea*

[2] *Division of Environmental Science and Engineering, Pohang University of Science and Technology, Pohang, South Korea*

[3] *School of Earth and Environmental Sciences, Seoul National University, Seoul, South Korea*

## Contents of this file

Figure S1 to S4

17 ## Supplementary

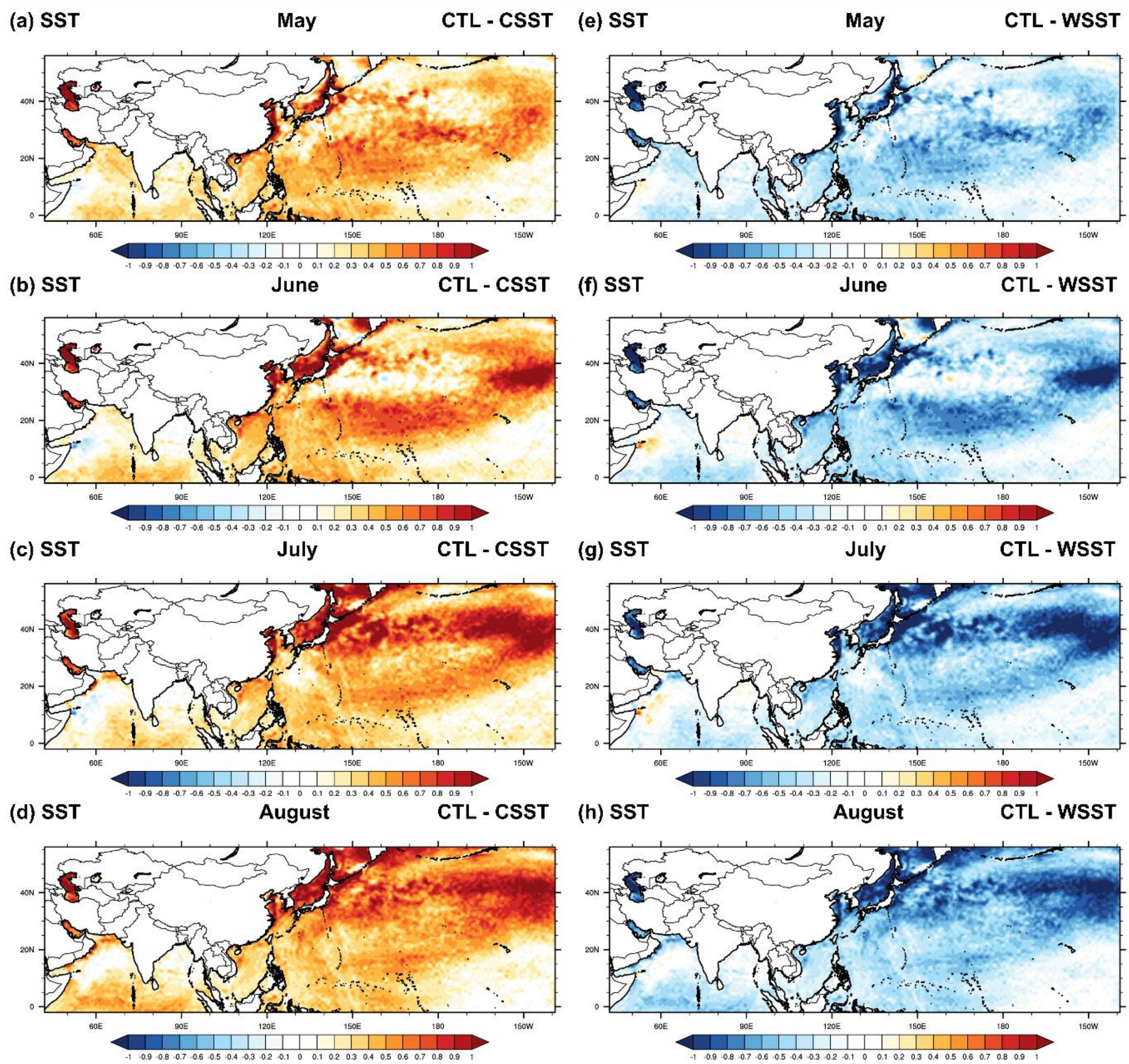

18

19 **Figure S1.** Differences in prescribed sea surface temperature between CTL and CSST (left
20 panel) and CTL and WSST (right panel). Each row from top to bottom shows the average
21 differences during May, June, July, and August. The plotted area also defines the model domain.

22

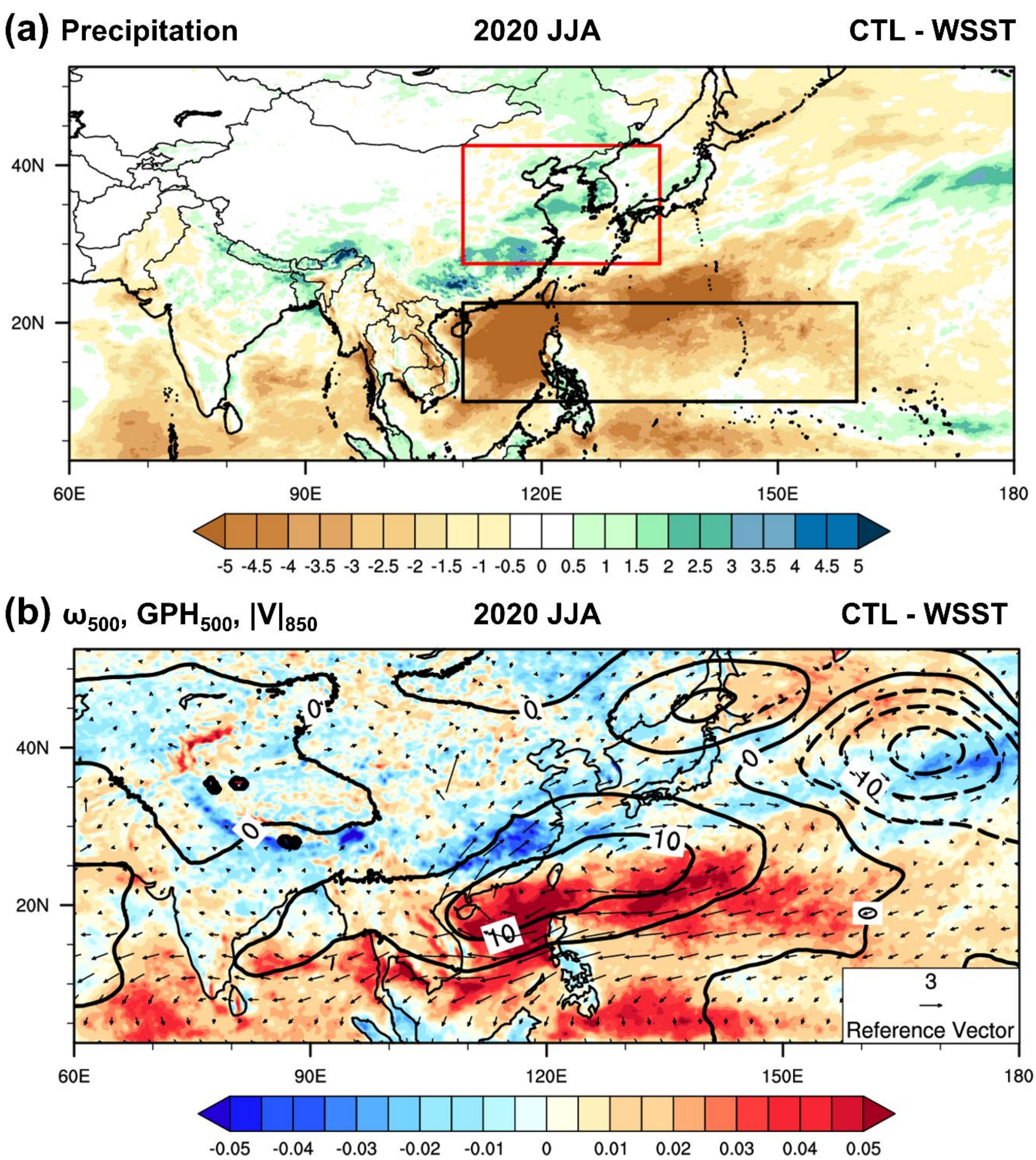

23

24 **Figure S2.** Differences of (a) daily mean precipitation (mm day$^{-1}$), and (b) 500 hPa geopotential
25 height (m; contour) and omega (Pa s$^{-1}$; shading), and 850 hPa wind vector (m s$^{-1}$; vector)
26 between CTL and WSST ensemble mean during 2020 JJA.

27

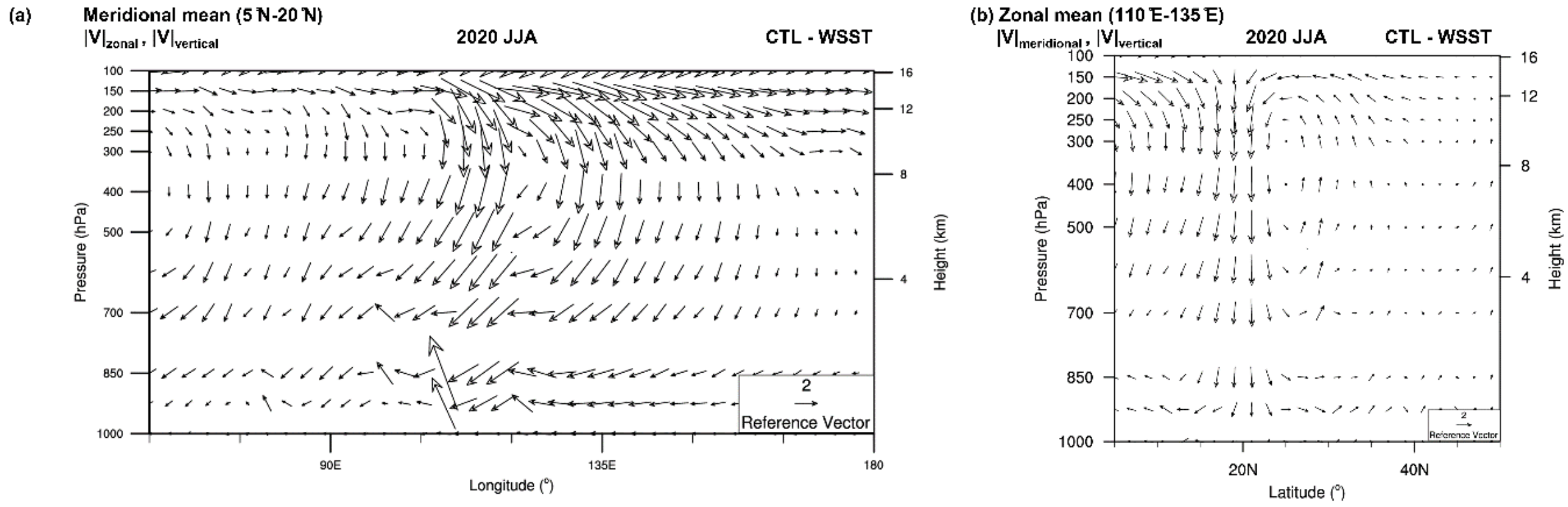

28

29 **Figure S3.** Differences of (a) the meridional mean (5°N-20°N) zonal wind (m $s^{-1}$) and vertical
30 velocity ($10^2$ m $s^{-1}$) and (b) the zonal mean (110°E-135°E) meridional wind (m $s^{-1}$) and vertical
31 velocity ($10^2$ m $s^{-1}$) between CTL and WSST ensemble mean during 2020 JJA.

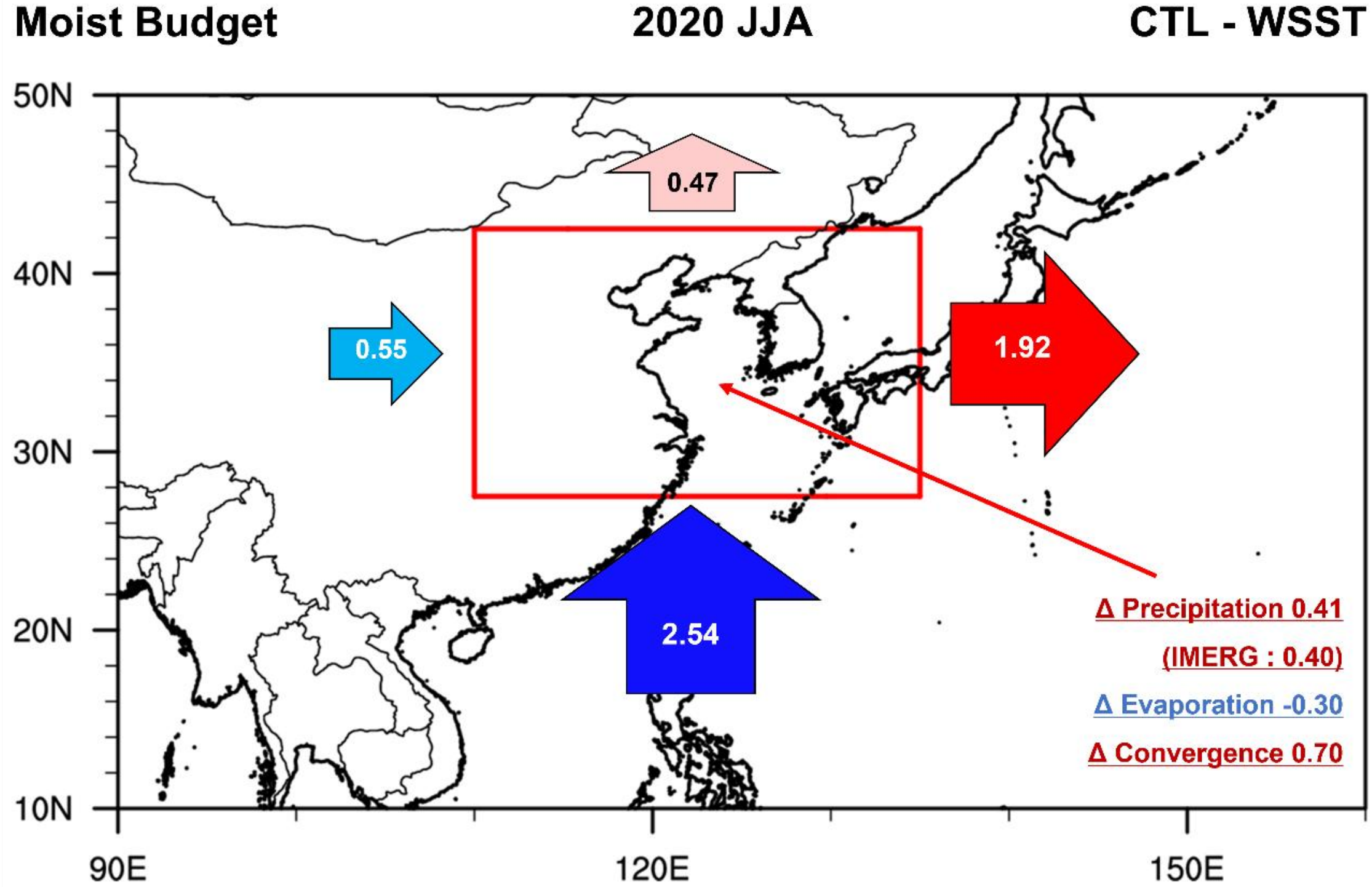

32

33 **Figure S4.** The moist budget for EASM of differences between CTL and WSST through each
34 boundary. The numbers in the blue arrows indicate the amount of water vapor (mm day$^{-1}$) that
35 is convergence into the EASM region at each boundary, and red, vice versa.